\documentstyle[prd,epsfig,aps,floats]{revtex}

\newcommand{\Sslash}[1]{ \parbox[b]{0.6em}{$#1$} \hspace{-0.55em}
                            \parbox[b]{0.55em}{ \raisebox{-0.2ex}{$/$}}}
\newcommand{\beq}{\begin{equation}}
\newcommand{\eeq}{\end{equation}}
\newcommand{\beqa}{\begin{eqnarray}}
\newcommand{\eeqa}{\end{eqnarray}}

\newcommand{\lsim}{\lesssim}

\newcommand{\cf}{{\it cf.}}
\newcommand{\etal}{{\it et al.\ }}  

\newcommand{\re}{{\rm Re}}
\newcommand{\tr}{{\rm Tr}}
\newcommand{\M}{{\cal M}}

\newcommand{\R}{{\cal R}}

\newcommand{\jpsi}{J/\psi}
\newcommand{\order}[1]{${\cal O}(#1)$}

\newcommand{\eq}[1]{Eq.\ (\ref{#1})}

\newcommand{\llb}{{\lambda\bar\lambda}}
\newcommand{\ssb}{{\sigma\bar\sigma}}

\newcommand{\avec}{\bbox{a}}
\newcommand{\betavec}{\bbox{\beta}}
\newcommand{\ovec}{\bbox{0}}
\newcommand{\pvec}{\bbox{p}}

\newcommand{\Gvec}{\bbox{G}}
\newcommand{\Hvec}{{{\rm \bf{H}}}}
\newcommand{\Gammavec}{\bbox{\Gamma}}

\newcommand{\qvec}{\bbox{q}}

\newcommand{\evec}{\bbox{e}}
\newcommand{\lvec}{\bbox{\ell}}
\newcommand{\sigmavec}{\bbox{\sigma}}

\newcommand{\evechat}{{\hat e}}

\newcommand{\avechat}{{\hat a}}
\newcommand{\Gvechat}{{\hat G}}
\newcommand{\pvechat}{{\hat p}}

\newcommand{\as}{\alpha_s}

\newcommand{\ket}[1]{\vert{#1}\rangle}

\newcommand{\qpair}{Q\bar Q}

\newcommand{\PL}[3]{Phys.\ Lett.\ {{\bf#1}}, {#2} ({#3})}
\newcommand{\NP}[3]{Nucl.\ Phys.\ {{\bf#1}}, {#2} ({#3})}
\newcommand{\PR}[3]{Phys.\ Rev.\  {{\bf#1}}, {#2} ({#3})}
\newcommand{\PRL}[3]{Phys.\ Rev.\ Lett.\ {{\bf#1}}, {#2} ({#3})}
\newcommand{\ZP}[3]{Z. Phys.\ {{\bf#1}}, {#2} ({#3})}

\begin{document}

\twocolumn[\hsize\textwidth\columnwidth\hsize\csname
@twocolumnfalse\endcsname
\title{%
\hbox to\hsize{\normalsize\hfil\rm NORDITA-2000/43 HE}
\hbox to\hsize{\normalsize\hfil\rm LAPTH-793/00}
\hbox to\hsize{\normalsize\hfil hep-ph/0004234}
\hbox to\hsize{\normalsize\hfil April 26, 2000} 
\vskip 40pt
Quarkonium Production through Hard Comover Scattering. II
\cite{byline1}}
\author{Nils Marchal and St{\'e}phane Peign{\'e}}
\address{LAPTH/LAPP, wwwlapp.in2p3.fr\\
F-74941 Annecy-le-Vieux Cedex, France\\}
\author{Paul Hoyer}
\address{Nordita, www.nordita.dk\\
Blegdamsvej 17, DK--2100 Copenhagen, Denmark\\}
\maketitle

\begin{abstract}
We extend to large transverse momentum $p_{\perp}$  an approach to
quarkonium hadroproduction previously suggested for low
$p_{\perp}$. The dynamics involves a perturbative rescattering
of the heavy quark pair off a comoving color field which originates from
gluon radiation prior to the heavy quark production vertex. Assuming
simple properties of the comoving field we find the rescattering scenario
to be in reasonable agreement with data. At large $p_{\perp}$,
$\psi'$ is predicted to be unpolarized, 
and $\chi_1$ production is favoured compared to $\chi_2$. 
We predict the $\chi_1$ polarization to
be transverse at low $p_{\perp}$, and to get a longitudinal component
at large $p_{\perp}$.
\end{abstract}
\vskip2.0pc]

\section{Introduction} \label{sec1}

Quarkonium production is a rich domain of QCD which involves several
hardness scales, namely
the transverse momentum $p_{\perp}$ of the
heavy quark pair, the quark mass $m$ and the Bohr momentum
of the heavy quarks in quarkonium $\as m$. Furthermore,
bremsstrahlung gluons are produced along with the pair at an intermediate
hardness scale.
The proximity of the open flavour
threshold makes the production of a quarkonium state highly sensitive
to its production environment.
Relatively soft
rescatterings of the quark pair with surrounding fields can make or
break the bound state.

Quarkonium production may be contrasted with open heavy flavour production,
for which the total cross section is unaffected by late rescattering. This
insensitivity is technically expressed through the QCD factorization
theorem \cite{fact}, which allows the cross section to be written in terms of
process-independent parton distributions. An analogous factorization of
hard and soft physics does not apply to quarkonium production rates,
which constitute a small
fraction of the total open flavour cross section. This makes
quarkonium physics a challenging and potentially rewarding field which can
teach us new aspects of the dynamics of hard processes.

Attempts to ignore rescattering effects in quarkonium production have met
with success in photoproduction but failed in hadroproduction. The
PQCD prediction for quarkonium production without rescattering,
at lowest order in the quark pair relative velocity $v$
(the `Color Singlet Model',
CSM \cite{csm,kzsz}), 
accounts well for the HERA data on $\gamma p \to \jpsi X$
in the photon fragmentation region \cite{photohera}.
On the contrary, the CSM underestimates the Tevatron $\bar p p$ 
data on direct $\jpsi$ and $\psi'$ 
production at large $p_{\perp}$
by a factor $\sim 50$ \cite{psiprimeanomaly}. 
Similar discrepancies are also observed in hadroproduction
at low $p_{\perp}$ \cite{rev3}.

The CSM amplitudes contain infrared divergencies, which for $P$-wave states
appear already at lowest order in $\as$. The divergencies cancel
systematically between terms of different orders in $v$,
as shown by the `NRQCD' \cite{nrqcd} expansion of the
amplitude. In the absence of rescattering effects NRQCD is thus a
theoretically viable framework for calculating quarkonium production in
QCD.

The NRQCD terms of higher order in $v$ correspond to relativistic effects
in the quarkonium wave function, and are related to its higher Fock
components such as $\ket{Q\bar Q g}$.
The magnitude of the relativistic corrections
is poorly known, and higher powers of $v$ may be compensated by lower
powers of $\as$ at the hard vertex.
Hence one may consider the possibility that
the observed discrepancies between the
CSM and hadroproduction data
are due to higher order contributions in the NRQCD expansion
(the `Color Octet Model', COM \cite{com}).
This implies, however, new contributions also in photoproduction which
tends to lead to an overestimate of the data \cite{photohera}.
The phenomenological difficulties of a COM approach \cite{hp,paulstalk}
have recently been made more acute by the (preliminary)
CDF data on $\psi'$ polarization at large $p_\perp$ \cite{cdf99}.
The COM prediction of a transverse $\psi '$ polarization,
considered as a crucial test of the COM approach,
fails by more than 3 standard deviations \cite{bkl}.

It would thus appear that rescattering is
important in
quarkonium hadroproduction. Since gluons, unlike
photons,
carry a comoving color field one expects more rescattering in
hadroproduction than in photoproduction. In \cite{hp} (referred to as I in
the following) a scenario involving a perturbative rescattering of
$\qpair$ pairs produced at low $p_\perp \lsim m$ was studied.
A simple form of the comoving field
allowed several of the puzzling features of the data to be understood.

Here we shall extend the rescattering picture of I to quarkonium
production at high $p_\perp$. We refer to I for a detailed discussion of
the rescattering mechanism and its phenomenological justification.
In section \ref{sec2} we present the low $p_\perp$ calculation in
a more compact form than in I and recall the results which were obtained.
In section \ref{sec3} we propose a generalization of the low
$p_\perp$ approach to high $p_\perp$, and evaluate $S$-wave and $P$-wave
quarkonium
production amplitudes.  All results obtained within the rescattering
picture (at low and high $p_\perp$) are summarized in section \ref{sec4},
together with a discussion of our assumptions. 

\section{Low $p_\perp$ Quarkonium Production} \label{sec2}

\subsection{Physical picture} \label{sec21}

The rescattering mechanism of quarkonium hadroproduction proposed in
I is based on an interaction of the heavy quark pair with  a comoving color
field. Let us recall why a color field
comoving with the $\qpair$ pair may be created in hadroproduction, but not
in photoproduction.

In the QED process  $e^+e^- \to \mu^+\mu^-$ near threshold (Fig.~1a), the
incoming electrons are surrounded by an electromagnetic (photon)
field. As the electrons annihilate, these fields pass through each other
without interacting (except via \order{\alpha_{em}} processes). The
$\mu^+\mu^-$ pair is thus created in a field-free environment.

In the analogous QCD process $gg\to \qpair$ (Fig.~1b), the
color fields associated with the color charge of the colliding gluons on
the contrary interact strongly. Such multiple interactions may create a remnant
color field  with low rapidity components in the $\qpair$ rest frame.
Rescattering may 
occur between this remnant field and the $\qpair$ pair.

In photoproduction, $\gamma g\to \qpair$ (Fig.~1c), the (unresolved)
photon carries no radiation field and the situation is analogous to
QED:
no remnant field is comoving with the $\qpair$ pair.
Rescattering thus occurs only in hadroproduction, and may explain some of
the observed `anomalies' of quarkonium production. It is absent in
photoproduction (except for resolved photon contributions), where the
CSM predictions in fact are quite satisfactory.
\begin{figure}[htb]
\center\leavevmode
\epsfxsize=6cm
\epsfbox{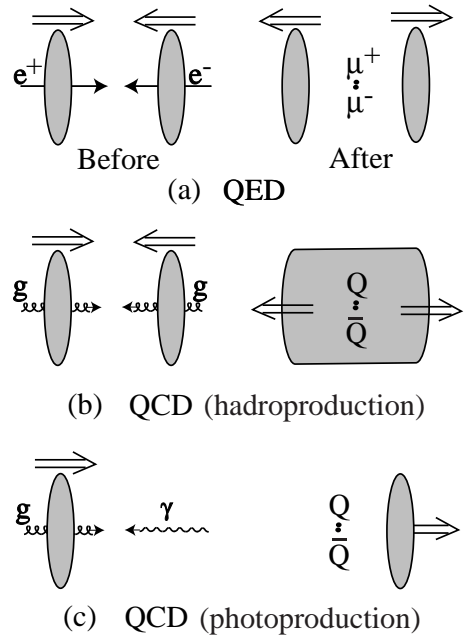}
\medskip
\caption[dummy]{Scenario
for the creation of a comoving color
field in low $p_\perp$ hadroproduction.}
\end{figure}

\subsection{The model} \label{sec22}

The effects on low $p_\perp$ quarkonium hadroproduction
of a rescattering of the
$\qpair$ pair off a comoving color field
were
studied in
I in the framework of a simple model. Before extending this scenario to
quarkonium production at high $p_\perp$ we summarize the main results of I.

The model applies to high energy production, where the quarkonium
carries a moderate fraction $x_F$ of the hadron beam momentum and has
$p_\perp \sim m$, where $m$ is the heavy quark mass. The amplitude for
quarkonium production in the bound state rest frame shown in
Fig.~2,
\beqa
\M &=& \sum_{L_z,S_z} \langle LL_z; SS_z |JJ_z \rangle
\sum_{\llb,\ssb} \int \frac{d^3\pvec}{(2\pi)^3} \frac{d^3\pvec '}{(2\pi)^3}
d^3\qvec \nonumber \\ &\times&
\delta^3(\pvec + \pvec ' + \lvec) \Phi_\llb^{[8]}(\pvec, \pvec ')\
\R_{\llb,\ssb}(\lvec, \pvec,\qvec)\ {\Psi_\ssb^{L_zS_z}(\qvec)}^*
\nonumber \\
\label{prodampl}
\eeqa
is a convolution of the $gg \to \qpair$ color octet
wave function $\Phi^{[8]}$, the kernel $\R$ describing
the rescattering between the pair and the
comoving color field, and the quarkonium wave function $\Psi$.

\begin{figure}[htb]
\center\leavevmode
\epsfxsize=6cm
\epsfbox{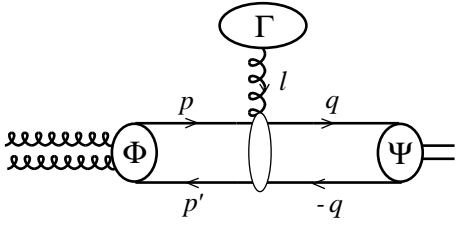}
\medskip
\caption[dummy]{The low $p_\perp$ quarkonium production amplitude induced by
     a rescattering on the comoving field $\Gamma$ (two diagrams). From
Ref.\cite{hp}.}
\end{figure}

The $gg\to \qpair$ process occurs on a proper time scale of order $m^{-1}$,
whereas the time scale for quarkonium bound state formation is of
order $(\alpha_s m)^{-1}$. The rescattering is assumed to be
characterized by the `semi-hard' scale $\mu$ of DGLAP
gluon radiation
\begin{equation}
\alpha_s m \ll \mu \ll m
\label{mu}
\end{equation}
As a consequence, the rescattering is well separated in time from both
the $\qpair$ pair creation and the bound state formation. The heavy
quarks thus propagate nearly on-shell both before and after the
rescattering.
In \eq{prodampl} the helicities and momenta of the quarks (in the
quarkonium rest frame) before the rescattering are denoted $\lambda, \bar
\lambda$ and $\pvec, \pvec '$. After the rescattering of momentum transfer
$\lvec$ the same quantities are denoted by $\sigma, \bar \sigma$ and
$\qvec, -\qvec$. The calculation of I can be summarized by expressing
every factor of  (\ref{prodampl}) as a contraction between the
spinors
$\chi_\lambda$ and $\chi_\lambda^\dagger$, with
$\chi_+^\dagger = \left( 1\ 0\right)$,
$\chi_-^\dagger = \left( 0\ 1 \right)$,

\begin{mathletters} \label{convfactors}
\beqa
\Phi_\llb^{[8]}(\pvec, \pvec ') &=& \chi_{\lambda}^\dagger \, \tilde
\Phi^{[8]}(\pvec, \pvec ') \, \chi_{-\bar\lambda} \label{phi} \\
\R_{\llb,\ssb}(\lvec,\pvec,\qvec) &=&
    (2\pi)^3 \delta^3(\pvec -\qvec
+\lvec) \, \delta_{\bar\lambda}^{\bar\sigma} \,
\chi_{\sigma}^\dagger \, \tilde \R^Q(\lvec,\qvec) \, \chi_{\lambda}
\nonumber   \\
+ &(2\pi)^3& \delta^3(\pvec -\qvec)
\,\delta_{\lambda}^{\sigma}
\, \chi_{-\bar \lambda}^\dagger \, \tilde \R^{\bar Q}(\lvec,-\qvec) \,
\chi_{-\bar \sigma}  \label{r} \\
{\Psi_\ssb^{L_zS_z}(\qvec)}^* &=& \frac{\Psi_{LL_z}^* (\qvec)}{\sqrt{2}}
\, \chi_{-\bar\sigma}^\dagger \, \evechat (S_z)^* \, \chi_\sigma
\label{psi}
\eeqa
\end{mathletters}
In $\tilde \R^Q(\lvec,\qvec)$ and $\tilde \R^{\bar Q}(\lvec,\qvec)$,
$\qvec$ stands for
the momentum of the scattered quark or antiquark {\it after} the
scattering. \eq{psi} holds for $S=1$ quarkonia, $\evec (S_z)$ being the
bound state spin polarization vector defined in the rest frame as
\begin{mathletters} \label{polvec}
\beqa
\evec(\pm 1) &=& (\mp 1,-i,0)/\sqrt{2} \label{evectrans}\\
\evec(0) &=& (0,0,1) \label{eveclong}
\eeqa
\end{mathletters}
We use the notation $\avechat= \avec \cdot \sigmavec$ for a vector
$\avec$, with $\sigma_k$ the Pauli matrices.
The amplitude (\ref{prodampl}) takes a form which can be readily inferred
from Fig.~2,
\beqa \label{spinoneprodampl}
\M(S=1)&=& \sum_{L_z,S_z} \langle LL_z; SS_z |JJ_z \rangle
\int \frac{d^3\qvec}{(2\pi)^3} \frac{\Psi_{LL_z}^* (\qvec)}{\sqrt{2}}
\nonumber \\ \frac{1}{\sqrt{3}} T^b_{ji}
& &\left\{\tr \left[\evechat (S_z)^* \tilde
\R^Q(\lvec,\qvec) \tilde \Phi^{[8]}(\qvec-\lvec, -\qvec)\right] \right.
\nonumber \\ &+&  \left. \tr \left[\tilde \R^{\bar Q}(\lvec,-\qvec)
\evechat (S_z)^* \tilde \Phi^{[8]}(\qvec, -\qvec -\lvec) \right]
\right\}
\eeqa
where $\tr$ denotes the trace of $2 \times 2$ matrices.
For convenience we separate the rescattering color factor from the
kernel $\R$. The color indices of the rescattering gluon and of the quark
and antiquark before the rescattering are denoted by $b$, $i$, $j$.
The general form (\ref{spinoneprodampl}) will also be used in the case of
high $p_\perp$ quarkonium production via gluon fragmentation in the
next section.

In \eq{spinoneprodampl} the zeroth and first order terms in $\qvec$
correspond to $S$ and $P$-wave production, respectively.
Recalling that $|\lvec| \sim \mu \ll m$ we work only to first
order in the small quantities $\lvec/m$, $\qvec/m$. The
rescattering kernel then reads \cite{hp}
\begin{mathletters} \label{rescattkernel}
\beqa
\tilde \R^{Q, \bar Q}(\lvec,\qvec) &=& \frac{g}{2m}
\left[ \mp i K + \Gvechat \right] \label{rq}\\
K(\lvec,\qvec) &=& 2m \Gamma^0(\ell^0, \lvec) +
\Gammavec(\ell^0, \lvec) \cdot (\lvec -2
\qvec) \label{K}\\
\Gvec(\lvec) &=& \Gammavec(\ell^0, \lvec) \times \lvec  \label{Gvec}
\eeqa
\end{mathletters}
The color field comoving with the $\qpair$ pair is modelled as
an external source $\Gamma^{\mu}(\ell)$, the form of which is
not known {\em a priori}. The hardness scale $\mu$ of $\Gamma$
is given by \eq{mu}.

The wave function $\tilde \Phi^{[8]}$ of the quark pair produced in $gg
\to \qpair$ is \cite{hp}
\beqa \label{phiampl}
\tilde \Phi^{[8]}(&\pvec&, \pvec ') = g^2 T^c_{ij}
\left\{ i \left(d_{a_1 a_2 c}+
\frac{\delta \pvec^z}{2m}if_{a_1 a_2 c}\right) (\evec_1 \times \evec_2)^z
\right. \nonumber\\
&+&\left. \frac{d_{a_1 a_2 c}}{2m} \left[ \evec_1 \cdot
\evec_2\,\, \delta \pvec^z \sigma_3 + \evec_1 \cdot \delta \pvec\,\,\evechat_2
+\evec_2 \cdot \delta \pvec\,\, \evechat_1  \right] \right\} \nonumber
\\ \label{phioctet}
\eeqa
with $\delta \pvec = \pvec - \pvec '$, $a_1$, $a_2$ the color indices
of the incoming gluons and $\evec_i = \evec(\lambda_i)$,
$i=1, 2$ their polarization vectors.
Using (\ref{spinoneprodampl}), (\ref{rescattkernel}) and (\ref{phiampl}) we
recover Eq.~(18) of I for $^3S_1$ production\footnote{In Eq.~(18) of
I, the gluon propagator $i/\lvec^2$ should be included in
$\Gamma^{\mu}$ and the color factor $D^d$ should be multiplied by
$\delta_i^{i'} \delta_j^{j'}$.
These details are not important for
the conclusions of I.}
\beqa
\M(&^3S_1&,S_z) = \frac{d_{a_1 a_2 b}}{4\sqrt{3}}
\frac{2g^3R_0}{\sqrt{2\pi m^3}}
\left\{ i\lambda_1 \delta_{\lambda_1}^{-\lambda_2}
\Gvec \cdot \evec(S_z)^* \right. \nonumber \\
-\left. \right.&\Gamma^0&(\ell)\left. \left[\delta_{\lambda_1}^{-\lambda_2}
\lvec^z \, \delta_{S_z}^0 - \evec_1 \cdot \lvec \,
\delta_{S_z}^{\lambda_2} - \evec_2 \cdot \lvec \,
\delta_{S_z}^{\lambda_1} \right] \right\}
\label{swaveampl}
\eeqa
Here we used for $S$-wave bound states
\beq
\int \frac{d^3\qvec}{(2\pi)^3} \Psi_{00}^*(\qvec) =
\frac{R_0}{\sqrt{4\pi m}}
\label{swavenorm}
\eeq
where $R_0$ is the wave function at the origin.
Summing over $a_1$, $a_2$, $b$ and $\lambda_1$, $\lambda_2$ yields

\beqa
&\sum& |\M(^3S_1,S_z)|^2 =
\frac{(N_c^2-4)(N_c^2-1)}{16N_c} \frac{4 g^6 R_0^2}{3 \pi m^3} \nonumber
\\
&\ & \times \left\{
\begin{array}{cc}
{\left|\Gvec^z \right|}^2
+ (\lvec^z)^2 {\left| \Gamma^0(\ell) \right|}^2  & (S_z=0) \\ & \\
\begin{array}{l}
\frac{1}{2}{\left|\Gvec_\perp \right|}^2
+ \frac{3}{2}(\lvec_\perp)^2 {\left| \Gamma^0(\ell) \right|}^2  \\
- \re \left[i \Gvec^y \Gvec^{x*}  \right]
\end{array}
    & (S_z=+1) \\
\end{array} \right.
\label{samplsq}
\eeqa
where no assumption has so far been made on $\Gamma^{\mu}$.

\subsection{Results and predictions}
\label{sec23}

$S$-wave states are observed to be unpolarized at low $p_\perp$
\cite{badier,poln,heinrich}. Comparing to \eq{samplsq}, this suggests
that the gluons in $\Gamma$ are dominantly transverse ($|\Gamma^0| \ll
|\Gammavec|$) and isotropically distributed. Hence we use the ansatz
\beq
\Gamma^{\mu}(\ell) \rightarrow e^{\mu}_{\lambda}(\ell) \
{\rm with} \ \lambda = \pm 1
\label{ansatz}
\eeq
where $e^\mu_{\pm 1}(\ell)$
denotes a transverse
polarization vector for the gluon $\ell$.
The isotropy assumption enters in the integral over $\lvec$ after the sum
over $\lambda$ is performed.
In (\ref{samplsq}) we obtain
\beqa \label{isotropy}
\int \sum_{\lambda} {\left|\Gvec^z \right|}^2 &\rightarrow& \int \left[
(\lvec^x)^2  +  (\lvec^y)^2 \right]
\rightarrow \frac{2}{3} \int \lvec^2 \nonumber \\
\int \sum_{\lambda} \frac{1}{2} {\left|\Gvec_\perp \right|}^2
&\rightarrow&
\int \left[ \frac{(\lvec^x)^2 + (\lvec^y)^2}{2}  +  (\lvec^z)^2 \right]
\rightarrow \frac{2}{3} \int \lvec^2 \nonumber \\
& & \int \sum_{\lambda} \Gvec^y \Gvec^{x*} \rightarrow 0
\eeqa
The treatment of $P$-wave production is similar. For low $p_\perp$
production the results can be summarized as follows:

\begin{itemize}
\item Direct $S$-wave production is unpolarized.
\item P-wave states are produced in the ratio
$\sigma_{dir}(\chi_1) / \sigma_{dir}(\chi_2) = 3/5$.
\item Direct $\chi_1$ production is transversely polarized.
\item $\chi_2$ is produced only with $J_z=0$ and $J_z=\pm 1$
in the ratio
$\sigma_{dir}(\chi_2, J_z=0) / \sigma_{dir}(\chi_2, J_z=\pm 1) = 2/3$.
\end{itemize}

As noted in I, the decays of $^3P_J$ states produced via the rescattering
mechanism induce a
longitudinal $\jpsi$ polarization. Since the
$\jpsi$ is observed to be nearly unpolarized \cite{badier,poln}, this
indicates that also the CSM mechanism contributes to $\chi_2$ production.
CSM produced $\chi_2$ states have $J_z=\pm 2$ only and
decay to transversely polarized $\jpsi$'s. For
$\sigma_{CSM}(\chi_2) \simeq \sigma_{rescatt}(\chi_2)$, an overall
$\jpsi$ polarization consistent with the data is obtained. Hence we
also get a lower $\chi_1 / \chi_2$ ratio,
$\sigma_{dir}(\chi_1) / \sigma_{dir}(\chi_2) \simeq 0.3$,
compatible with the data \cite{alexopoulos}.

It is interesting to observe that (low $p_\perp$) $P$-wave production via
rescattering is actually independent of $\Gamma^0$. This
is because only the $S=0$, $L=1$ part of the $\qpair$ wave function
(\ref{phiampl}) (the term proportional to $f_{a_1 a_2 c}$) contributes
in this case.
Thus a spin-flip rescattering (transverse gluon exchange) is required to
form a $^3P_J$ state.

\section{Quarkonium production at high $p_\perp$}
\label{sec3}

\subsection{Scenario for rescattering} \label{sec31}

Quarkonium production at high $p_\perp \gg m$ proceeds dominantly through
the fragmentation of quasi-transverse gluons \cite{fragmentation}.
DGLAP radiation from the fragmenting gluon
gives rise to a color field comoving with the quark pair, and we
assume that rescatterings occur between this field and the pair.
The field comoving with a high $p_\perp$ $\qpair$ pair
is {\em a priori}
different from the one at low $p_\perp$.
\begin{figure}[htb]
\center\leavevmode
\epsfxsize=6cm
\epsfbox{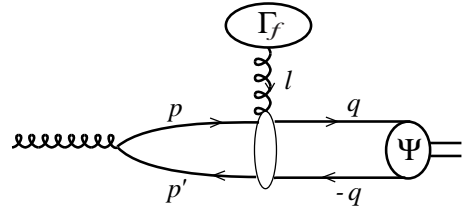}
\medskip
\caption{High $p_\perp$ quarkonium production via gluon fragmentation.
     The $\qpair$ pair rescatters on the comoving
     color field $\Gamma_f$ before forming quarkonium.}
\end{figure}

A schematic view of our model is shown in Fig. 3 for $^3P_J$ quarkonia
(which require a single rescattering). A high $p_\perp$ transverse gluon
(typically produced in a process like $gg \to gg$, which does not concern us
here) fragments into a heavy $\qpair$ pair. The pair rescatters from an
external field $\Gamma_f$ of hardness $\mu_f$ similar to the scale $\mu$ of
\eq{mu} and forms a quarkonium bound state. The amplitude of the
fragmentation process is a convolution, analogous to \eq{prodampl}, of the
quark pair wave function $\Phi_{f,\llb}^{[8]}$, the rescattering kernel
$\R_{\llb,\ssb}(\lvec, \pvec,\qvec)$ given by Eqs. (\ref{r}) and
(\ref{rescattkernel}), and the quarkonium wave function
$\Psi_\ssb^{L_zS_z}(\qvec)$ of \eq{psi}. As in the low $p_\perp$
situation, the heavy quarks are assumed to be on-shell before and after
the rescattering.

The wave function $\Phi_{f, \llb}^{[8]}$ of the pair in the quarkonium rest
frame is
\begin{mathletters} \label{phifrag}
\beqa
&\Phi_{f, \llb}^{[8]} = \chi_{\lambda}^\dagger \, \tilde
\Phi_f^{[8]} \, \chi_{-\bar\lambda} = g \, T^a_{ij} \,
{\bar u}_{\lambda}(\pvec) \Sslash{e}(\lambda_g) {v}_{\bar
     \lambda}(\pvec ')    \label{phifdef} \\
&\tilde \Phi_f^{[8]}(\pvec, \pvec ') =  -2m g \, T^a_{ij}
\left\{\evechat(\lambda_g) + \frac{\pvechat}{2m}\,
\evechat(\lambda_g)
\, \frac{\pvechat '}{2m}\right\}  \label{phif}
\eeqa
\end{mathletters}
Here $a$, $i$, $j$ are the color indices of the gluon and quarks and
$\evec(\lambda_g)$ is the polarization vector of the fragmenting gluon.
For the range $p_\perp \gg m$ that we are considering the gluon is
quasi-real and hence transversely polarized ($\lambda_g =\pm 1$) relative
to its direction in the laboratory frame.
An expansion in the small ratios $\pvec/m$,
$\pvec'/m$ has been made in \eq{phif}.

\subsection{Production of $^3S_1$ states }
\label{sec32}

Two rescatterings are needed to produce $^3S_1$ states
via gluon fragmentation, see Fig.~4. The fragmentation
amplitude $\M_{f}$ can thus be expressed
as a convolution similar to (\ref{spinoneprodampl}), with
two rescattering factors $\R$. Only the first term in the $\qpair$ wave
function (\ref{phif}), corresponding to a gluon fragmenting into a $\qpair$
pair with $L=0$, actually contributes to $S$-wave production.
\begin{figure}[htb]
\center\leavevmode
\epsfxsize=8cm
\epsfbox{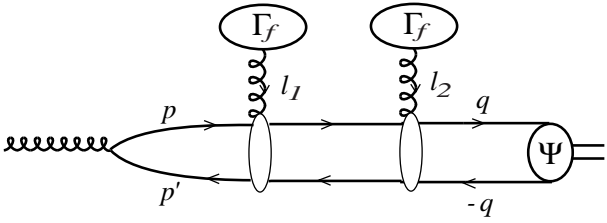}
\medskip
\caption{High $p_\perp$ $^3S_1$ production in the rescattering
     scenario (four Feynman diagrams).}
\end{figure}
\eject
We find for the $^3S_1$ production amplitude
\beqa \label{fragsampl}
& &M_f(^3S_1,S_z) = \frac{-mg R_0}{\sqrt{6\pi m}} \nonumber
\\ \times
\left\{\right.&\tr&\left. \left[a\, b_2\, b_1 \right] \tr
\left[\evechat (S_z)^*
\tilde \R^Q(\lvec_2,\ovec) \tilde \R^Q(\lvec_1,-\lvec_2)
\evechat (\lambda_g)\right] \right. \nonumber \\
+  \left.  \right.&\tr& \left. \left[a\, b_2\, b_1 \right] \tr
\left[\tilde
\R^{\bar Q}(\lvec_2,\ovec) \evechat (S_z)^* \tilde \R^Q(\lvec_1,\ovec)
\evechat (\lambda_g)\right] \right. \nonumber \\
+  \left.   \right.&\tr&\left. \left[a\, b_1\, b_2 \right] \tr
\left[\evechat (S_z)^*
\tilde \R^Q(\lvec_2,\ovec) \evechat (\lambda_g) \tilde
\R^{\bar Q}(\lvec_1,\ovec) \right] \right. \nonumber \\
+ \left. \right.&\tr&\left. \left[a\, b_1\, b_2 \right] \tr
\left[\tilde \R^{\bar Q}(\lvec_2,\ovec) \evechat (S_z)^* \evechat
(\lambda_g)
\tilde \R^{\bar Q}(\lvec_1, -\lvec_2) \right] \right\} \nonumber \\
\eeqa
where $b_1$, $b_2$ are the color
indices of the two rescattering gluons and
$\tr \left[a\, b_1\, b_2 \right]$ is a shorthand notation for
$\tr \left[T^a T^{b_1} T^{b_2} \right]$.
The four terms in (\ref{fragsampl}) correspond to the four Feynman
diagrams implied in Fig.~4. Using
(\ref{rescattkernel}) leads to the surprisingly simple result
\begin{equation}
\M_f(^3S_1,S_z) = \frac{- R_0 g^3}{2 \sqrt{6\pi m^3}} \,
d_{ab_1b_2} \Gvec_1 \cdot \evec(\lambda_g) \, \Gvec_2 \cdot \evec(S_z)^*
     \label{highptsampl}
\end{equation}
where $\Gvec_i = \Gammavec_f(\lvec_i) \times \lvec_i$. Note
that (\ref{highptsampl}) is independent of $\Gamma_f^0$. Only transverse
gluons in the comoving gluon field contribute to $^3S_1$ quarkonium
production at high $p_\perp$.

The expression
(\ref{highptsampl}) trivially satisfies the gauge invariance
requirement $\M_f(\Gamma_f^{\mu}(\ell_i) \to \ell_i^{\mu}) = 0$.
Since the quarks are treated as being on-shell, each Feynman diagram
contributing to the production amplitude separately satisfies current
conservation. However, in order to keep track of gauge invariance when
combining the different kernels $\cal{R}$, one must pay attention
to their $\ell^0$-dependence.
For instance the value of $\ell_1^0$ in
$\tilde \R^{Q, \bar Q}(\lvec_1,-\lvec_2)$ differs from that in
$\tilde \R^{Q, \bar Q}(\lvec_1, \ovec)$ by an amount
$\delta \ell_1^0 = - \lvec_1 \cdot \lvec_2 / m$.
Now $\Gamma_f^{\mu}(\ell_1^0, \lvec_1)$ is related to
$\Gamma_f^{\mu}(\ell_1^0 + \delta \ell_1^0, \lvec_1)$ through a boost of
velocity $\betavec = - \lvec_2 /m$, $|\betavec| \ll 1$.
One realizes that the vector
$\Gammavec_f(\ell^0, \lvec)$ (and thus $\Gvec$) is insensitive to this
boost (at the level of our approximation) whereas
the difference between
$\Gamma_f^{0}(\ell_1^0, \lvec_1)$ and
$\Gamma_f^{0}(\ell_1^0 + \delta \ell_1^0, \lvec_1)$ is such that we get
\beq
\label{gaugeinv}
\tilde \R^{Q, \bar Q}(\lvec_1,-\lvec_2) \simeq
\tilde \R^{Q, \bar Q}(\lvec_1, \ovec)
\eeq
which gives the simple form (\ref{highptsampl}) for $\M_f$.

In the $S$-wave production amplitude (\ref{highptsampl}) the
dependence in $\lambda_g$ and $S_z$ is {\it factorized}.
No correlation appears between the fragmenting gluon and quarkonium
polarizations. Consequently, assuming only that $\Gamma_f$ is
isotropically distributed in the (comoving) $\qpair$ pair rest frame we
predict that directly produced $^3S_1$ quarkonia are unpolarized at high
$p_\perp$. Contrary to the low $p_\perp$ situation, we do
not now need to assume the dominance of transverse gluons in $\Gamma_f$.
The preliminary CDF data \cite{cdf99} on high $p_\perp$ $\psi'$
production seems to prefer a longitudinal polarization, but
the statistics is insufficient for a definite conclusion.

The high $p_\perp$ cross section ratio
$\sigma(\psi')/\sigma(\jpsi)$ appears to be larger \cite{rev3} than the
nearly universal ratio measured in low $p_\perp$ hadroproduction
\cite{vhbt,lourenco,gksssv},
\beq \label{dipole}
\left. \frac{\sigma(\psi ')}{\sigma_{dir}(\jpsi)}
\right )_{p_\perp \gg m} \simeq 0.4 >
\left. \frac{\sigma(\psi ')}{\sigma_{dir}(\jpsi)} \right
     )_{p_\perp \leq m} \simeq 0.24
\eeq
This is in qualitative agreement with our result that the
rescattering gluons are transverse and thus couple to the spin of the
individual quarks. In low $p_\perp$ production the target gluon is
longitudinal (in the target rest frame) and its coupling is proportional to
the $\qpair$ color dipole size $r_\perp$. The convolution (\ref{prodampl})
then probes the bound state wave function in relatively larger
configurations, where the node in the $\psi'$ wave function tends to
decrease its contribution \cite{knnz,hp2}. This systematics carries over
to diffractive photoproduction, where two longitudinal gluon exchanges
give a factor $r_\perp^2$, and the corresponding ratio (\ref{dipole}) is
measured to be  $\simeq 0.15$ \cite{photo}.

\subsection{Production of $^3P_J$ states }
\label{sec33}

In the case of high $p_\perp$ $^3P_J$ production, only one
scattering is needed (see Fig.~3). The production amplitude is given
by the expression (\ref{spinoneprodampl})
with $\tilde \Phi^{[8]}$ replaced by $\tilde \Phi_{f}^{[8]}$ of
\eq{phifrag}. We expand the curly bracket of (\ref{spinoneprodampl}) at
$\qvec \to \ovec$ and use for $P$-wave states
\begin{mathletters} \label{pwavenorm}
\beqa
\int \frac{d^3\qvec}{(2\pi)^3} \Psi_{1L_z}^*(\qvec) &=& 0  \\
\int \frac{d^3\qvec}{(2\pi)^3}\, \qvec \Psi_{1L_z}^*(\qvec) &=&
i\sqrt{\frac{3}{4\pi m}}{R_1}' \evec(L_z)^*
\eeqa
\end{mathletters}
where ${R_1}'$ is the derivative of the $P$-wave function at the
origin. In the small $\qvec$ expansion the linear terms in
$\qvec$ cannot arise from the rescattering kernel
$\tilde \R$ because of the gauge invariance property mentioned in the
preceding subsection. Recalling (\ref{gaugeinv}) with
$\lvec_2$ replaced by $-\qvec$, one indeed sees that
$\tilde \R^{Q}(\lvec,\qvec)$ is actually independent
of $\qvec$. This means that the final orbital angular momentum of the $\qpair$
pair $L=1$ must be fixed before the scattering, via the $L=1$ part of
the $g \to \qpair$ wave function $\tilde \Phi_{f}^{[8]}$. This statement
remains true for any number of rescatterings. Thus only the second term of
(\ref{phif}) contributes to $P$-wave production. Inserting (\ref{rq}) and
(\ref{phif}) in (\ref{spinoneprodampl}) we arrive at
\begin{mathletters}
\beqa \label{pampl}
\M_f(^3P_J,J_z) &=& \frac{-g^2 R_1'/m}{4 \sqrt{2\pi m^3}} \delta^a_b
A^{\alpha \beta} B^{\alpha \beta, ij} \evec^i(\lambda_g) \Hvec^j \\
A^{\alpha \beta} &=&
\sum_{L_zS_z} \langle LL_z; SS_z |JJ_z \rangle \evec^{\alpha}(L_z)^*
\evec^{\beta}(S_z)^* \\
B^{\alpha \beta, ij} &=&
\delta_{i}^{j} \delta_{\alpha}^{\beta} + \delta_{i}^{\alpha}
     \delta_{j}^{\beta} - \delta_{i}^{\beta} \delta_{j}^{\alpha} \\
\Hvec &=& \lvec^2 \Gammavec_f + 2m \Gamma_f^0  \lvec \label{H}
\eeqa
\end{mathletters}
where $B^{\alpha \beta, ij}$ arises from the trace of four Pauli matrices.
Using the form of the tensor $A^{\alpha \beta}$ given in \cite{cholei}
we get
\beqa
\M_f(&^3P_J&,J_z) =  \frac{g^2 \sqrt{3}R_1'/m}{4 \sqrt{2\pi m^3}} \delta^a_b
\nonumber \\ &\times& \left\{
\begin{array}{cc}
\evec(\lambda_g) \cdot \Hvec & (J=0) \\
i\sqrt{\frac{2}{3}} \evec(\lambda_g)\cdot
(\Hvec \times \evec(J_z)^*) & (J=1) \\
0 & (J=2)
\end{array} \right.
\label{3pjampl}
\eeqa
The amplitude vanishes for $J=2$ since in this case $A^{\alpha \beta}$
is symmetric and $A^{\alpha}_{\alpha} = 0$. Experimentally
$\sigma(\chi_2) \simeq \sigma(\chi_1)$ at large $p_\perp$ \cite{papa}.
The cross section for $\chi_2$ production should be generated either by
more than one rescattering in our scenario or by
the CSM (or COM) mechanism of gluon emission.

To conclude this section we give our prediction for the $\chi_1$
polarization at large $p_\perp$ and discuss its dependence on the
rescattering field $\Gamma_f$.

The high $p_\perp$ fragmenting gluon being
quasi transverse we have from (\ref{3pjampl})
\beq
\label{chi1ampsq}
|\M_f^{\chi_1}(J_z)|^2 \propto  |(\Hvec \times \evec(J_z)^*)_{\perp}|^2
=  \left\{
\begin{array}{cc}
|\Hvec_{\perp}|^2 & (J_z=0) \\
|\Hvec^{z}|^2 & (J_z=\pm 1) \\
\end{array} \right.
\eeq
where the direction of the high $p_\perp$ gluon in the laboratory
frame is chosen as the reference axis.
For a general field $\Gamma_f$, the $\chi_1$ polarization depends on
the relative importance of the two terms contributing to the vector
$\Hvec$ given in (\ref{H}).
However, assuming that $\lvec$ is isotropically distributed,
we find the same result for the $\chi_1$ polarization parameter
\beq
\lambda_{\chi_1} = \frac{\sigma(J_z=+1)-\sigma(J_z=0)}
{\sigma(J_z=+1)+\sigma(J_z=0)} = -\frac{1}{3} \label{chi1pol}
\eeq
in the two following extreme cases:
\begin{itemize}
\item[(i)] $\Gamma_f$ contains only transverse gluons (the first term of
$\Hvec$ dominates) and is modelled by the ansatz (\ref{ansatz}) used
at low $p_\perp$.
\item[(ii)] $\Gamma_f$ is purely longitudinal (the second term of
$\Hvec$ dominates).
\end{itemize}
Thus, rather independently of the form of $\Gamma_f$, we predict the
$\chi_1$ yield to be longitudinally polarized if the rescattering
scenario dominates $\chi_1$ production at high $p_\perp$.

\section{Summary and discussion}
\label{sec4}
We presented an extension of the approach of I to high $p_\perp$
quarkonium production. As in our previous work, we assumed the
presence of rescattering between the heavy quark pair and a comoving
color field, and found that such a scenario qualitatively agrees with the
low {\it and} high $p_\perp$ data. We summarize here our main
results and predictions.

At low $p_\perp$, the rescattering mechanism applies only to hadroproduction,
not to photoproduction (in the photon fragmentation region). 
In order to agree with the
observed non-polarization of $^3S_1$ low $p_\perp$ hadroproduction, we
assume the field $\Gamma$ created at low $p_\perp$ to contain
dominantly transversely polarized gluons. The
$\sigma(\chi_1) / \sigma(\chi_2)$ ratio is found to be compatible with
the measured value and our main predictions at low $p_\perp$ are:
\begin{itemize}
\item[(i)] Direct $\chi_1$ production is transverse.
\item[(ii)] $\sigma_{dir}(\chi_2, J_z=0) / \sigma_{dir}(\chi_2, J_z=\pm 1)
= 2/3$. In addition, we expect a significant CSM (no rescattering)
contribution to $\sigma(\chi_2)$, having $J_z=\pm 2$.
\end{itemize}

The comoving field $\Gamma_f$ arising from high $p_\perp$ gluon
fragmentation is {\em a priori} different from the low $p_\perp$ field
$\Gamma$. Our scenario is identical in high $p_\perp$
hadroproduction and high $p_\perp$ photoproduction, the $\qpair$ pair
originating from gluon fragmentation in both cases. Our predictions at
large $p_\perp$ are:
\begin{itemize}
\item[(iii)] Direct $^3S_1$ production is unpolarized.
\item[(iv)] Direct $\chi_1$ production contains a longitudinal component
(\eq{chi1pol}).
\item[(v)] $\chi_2$ is not produced via (a single) rescattering.
\end{itemize}
The prediction (iii) is consistent with the present CDF data on $\psi '$
polarization \cite{cdf99}, but higher statistics is needed for a
definite conclusion. Only transverse gluon rescattering contributes to
$^3S_1$ production. This is in qualitative agreement
with the measured hadroproduction ratio $\sigma(\psi')/\sigma_{dir}(\jpsi)$
being smaller at low than at high $p_\perp$ 
(\cf\ \eq{dipole}), and being
even smaller
in diffractive photoproduction.

All our results and predictions apply equally to the charmonium and
bottomonium families.

We used several simplifying assumptions which we now discuss.
\begin{itemize}
\item We supposed the rescattering color field to be isotropically
distributed in the $\qpair$ rest frame. Our predictions on ratios and
relative polarization rates depend on this assumption.
\item We assumed the rescatterings to be perturbative. Thus only the
minimal number of rescatterings required to produce a given bound
state was considered. On the other hand the rescattering probability
should be large enough for this mechanism to dominate the CSM
$S$-wave and (low $p_\perp$)
$\chi_1$ rates. Whether such a compromise holds is
not obvious and will be studied in a future work.
\item It is unlikely that the production amplitude is sensitive only to the
quarkonium wave function (or its derivative) at the origin, at
least for charmonium. This is particularly so for $P$-waves
(see the discussion of $\sigma(\chi_2) / \sigma(\jpsi)$ in I) and when
`dipole' factors of $r_\perp$ enhance larger wave
function configurations \cite{hp2}.
\end{itemize}
\section*{Acknowledgments}
We are grateful for helpful discussions with T. Binoth, S. Brodsky, J.
Rathsman and U. Wiedemann.
N.M. would like to thank Nordita for its kind hospitality and support.

\end{document}